\documentclass{article}
\usepackage{epsfig,amssymb}

\def\beq{\begin{equation}}
\def\eeq{\end{equation}}
\def\bea{\begin{eqnarray}}
\def\eea{\end{eqnarray}}
\def\d{{\mathrm{d}}}

\newfont{\cursive}{pzcmi at 9pt}
\def\~t{\tilde{t}}

\begin{document}

\title{Compact binary coalescence parameter estimations for 2.5 post-Newtonian aligned spinning waveforms}
\author{Alex B. Nielsen \\ Max-Planck-Institute for
Gravitational Physics, \\ 
38 Callinstrasse, \\
30167 Hanover, \\
Germany}

\maketitle

\begin{abstract} 
We examine the parameter accuracy that can be achieved by advanced ground-based detectors for binary inspiralling black holes and neutron stars. We use the 2.5 PN spinning waveforms of Arun {\it et al.} \cite{arXiv:0810.5336}. Our main result is that the errors are noticeably different from earlier studies. An important contribution to this difference comes from self-spin terms at 2PN order not previously considered. While the masses can be determined more accurately, the individual spins are measured less accurately compared to previous work, in some cases by more than a factor of ten. We also examine several regions of parameter space relevant to expected sources and the impact of simple priors. A combination of the spins is measurable to higher accuracy and we examine what this can tell us about spinning systems.
\end{abstract}


\section{Introduction}

Upgrades to gravitational wave interferometric detectors are currently being performed \cite{Marx:2011hf}. Once completed these advanced detectors are expected to detect tens of astrophysical signals per year \cite{Abadie:2010cf}. The inspiral of massive compact bodies in binary systems is one of the most important expected signals. Detection of the gravitational waves from such events can provide information about the parameters of the compact bodies involved. Knowing the masses of the objects involved is central to classifying the observed objects as neutron star or black hole candidates. In addition, the objects may have intrinsic spin and General Relativity predicts that this will affect the motion of the bodies and the gravitational waveforms produced at a measurable level \cite{Apostolatos:1994mx}.

To detect the signals in the background noise the method of matched filtering is employed \cite{Allen:2005fk}. Theoretical templates based on Post-Newtonian (PN) expansions in General Relativity are employed to search the detector output for signals. The goodness of fit of a particular template to the data can be most easily seen by computing a signal-to-noise ratio (SNR). Because the detector is noisy, the best fitting set of parameters may not be the same as the true parameters of the source. The error in the estimation of the source parameters depends in part on the signal to noise ratio. A theoretical indication of the parameter errors can be computed by simulating signals and multiple realisations of the noise or by using the covariance matrix of Fisher information theory. We employ the Fisher information method here.

As the true waveform in General Relativity is approximated by a Post-Newtonian expansion, a systematic error is introduced when truncating the true waveform. From a theoretical viewpoint, it would be advantageous if this systematic error is smaller than other sources of noise.  The order to which the Post-Newtonian expansion is needed in order to provide a satisfactory model for detection and parameter extraction is still unknown, although there are some hints of at least a lower bound \cite{Cutler:1993vq} \cite{Tagoshi:1994sm}.

At the moment we do not possess PN expansions beyond the level that seem satisfactory and so it is prudent to use the highest order PN expansion available. Previous work has examined the measurement sensitivities of ground based interferometers for 1.5PN \cite{gr-qc/9402014} and 2PN \cite{gr-qc/9502040} \cite{Krolak:1995md} orders. Here we extend those works to the recently available 2.5PN spinning waveforms of Arun {\it et al.} \cite{arXiv:0810.5336} which also includes self-spin terms at 2PN not considered in previous work. Non-spinning waveforms are available to 3.5PN order and their parameter estimation has been examined in \cite{Arun:2004hn}.

As noted in \cite{Damour:1997ub} the gravitational wave flux for test particle non-spinning inspirals is badly behaved at 2.5PN level. Table I of \cite{Arun:2004hn} gives parameter estimation for non-spinning binaries at increasing steps of 0.5 in the PN expansion up to 3.5PN order. Although the convergence of the series looks very slow, the parameter errors for non-spinning waveforms shows no large fluctuations at the 2.5PN order.

We concentrate here on aligned spinning systems. With the assumption of aligned spin and orbital angular momentum axes, the waveforms depend on six parameters. Two of the parameters, $t_{c}$ and $\phi_{c}$, are extrinsic parameters determining the time and phase of the wave at coalescence. The point of coalescence, which is not well modeled by the PN expansion, could be replaced by any other chosen point in the inspiral - both parameters are effectively integration constants. Although the template waveform match does depend on their value, the parameter error ranges do not.

The other four intrinsic parameters are the masses and spins of the two inspiralling objects. It is ultimately only the masses and spins of the objects that we are interested in determining, but these parameters are correlated with $t_{c}$ and $\phi_{c}$. We will examine the effect of redefining the coalescence phase, $\phi_{c}$, and also the effect of imposing $2\pi$ periodic priors on this parameter.

Three important differences exist between the 2.5PN waveforms investigated here and earlier work for ground based detectors. Firstly the extension to 2.5PN order provides seven expansion coefficients in the waveform expanded as a series in the frequency, with only six parameters to be measured. These expansion coefficients are sometimes called ``chirp times`` in the time domain waveforms. Previous work at 1.5 PN order for ground based detectors \cite{gr-qc/9402014} considered five PN expansion coefficients depending on five parameters: $t_{c}$, $\phi_{c}$, the reduced mass, $\mu$, chirp mass, ${\cal{M}}$ and $\beta$, a parameter related to the spin.  Previous work at 2PN order \cite{gr-qc/9502040} considered six expansion coefficients with six parameters - adding an extra spin related parameter, $\sigma$. Adding more parameters typically reduces the accuracy at which any one parameter can be measured, but adding more expansion coefficients can counteract this effect. At 2.5PN order we are adding another coefficient without increasing the number of parameters, since the extra coefficient can be expressed as a combination of the others.

Secondly, the introduction of new terms independent of the frequency in the waveform phase greatly increases the correlation between the phase at coalescence, $\phi_{c}$ and the other parameters, in particular the spins. Errors in $\phi_{c}$ can then greatly magnify errors in the spins. This makes it even more imperative that errors in the $\phi_{c}$ are controlled in order not to over-contaminate the errors in the mass and spin parameters.

Thirdly, the waveform considered here contains self-spin interaction terms of the type spin(1)spin(1) and spin(2)spin(2) at 2PN order, that were not considered in previous analyses. The 2.5PN waveform of \cite{arXiv:0810.5336} contains these self-spin interaction terms at 2PN order, which are not present in the 2PN waveform used in \cite{gr-qc/9502040}. These terms were first calculated in \cite{Mikoczi:2005dn}. A major part of the increase in spin uncertainty can be traced to these self-spin interaction terms. The 2.5PN waveform used here, only contains spin-orbit terms at 2.5PN and it remains to be seen whether further, as yet uncalculated PN terms, will affect our results.

The first issue to be addressed in our work is; do the parameter errors differ significantly from results obtained using lower order PN expansions? The answer to this question will provide some guide to the question of what order of PN expansion is sufficient to model the true General Relativity signal to an acceptable level. A second question we will examine is; at this level of PN approximation and with the assumption of aligned spin and orbital angular moment, does intrinsic spin matter? To what extent can the experiments distinguish a spinning system from one without spin? A further, slightly different question is; how accurately can the expected data constrain the individual spins of each of the inspiralling objects?

For spinning black holes whose exterior is exactly described by the Kerr metric a theoretical upper limit $M^{2}\geq J$ exists bounding the magnitude of spin angular momentum, $J$, to be less than the square of the mass, $M$, in natural geometric units. Beyond this limit the object can no longer be described as a Kerr black hole but a naked singularity. This relationship can be expressed in terms of the spin parameter $\chi = J/M^2$ as $\chi\leq 1$. To truly claim that a massive object is a black hole one must show that the spin of the object satisfies this bound.

Black holes can be born with intrinsic spin or be spun up due to the accretion of matter. Theoretical studies indicate that the maximum amount of spin that a black hole can accumulate due to accretion corresponds to $\chi=0.998$ \cite{Thorne:1974ve}. A number of highly spinning black holes, with spin parameters $\chi > 0.9$ may have been found using X-ray techniques \cite{Gou:2011nq}.

Neutron stars are expected to have maximum spins beyond which they break up due to centrifugal forces \cite{Chakrabarty:2003kt}. The $\chi=1$ case for a neutron star corresponds to a frequency in the $10^4$Hz range. Ultra-fast millisecond pulsars have been observed with spins up to 700Hz. The three known binary pulsar systems that will coalesce in a Hubble time all have spins corresponding to $\chi\leq0.02$ \cite{Phinney:1991ei}. Non-compact objects, such as the Earth or Sun, actually do violate the $\chi=1$ bound, but it is not expected that such non-compact objects will be able to survive the tidal forces present in the last moments of inspiral in the frequency ranges of interest to us.

We will attempt a wider investigation of the 2.5 PN order parameter space by allowing the inspiralling objects to be spinning and examine the effect that this has on the parameter estimation. The errors in the parameters depend sensitively on the region of parameter space that is probed. Covering the full parameter space in detail is time-consuming and leads to the problem of presenting the resulting twelve dimensional space - the six parameter errors for the six parameters. We will instead pick a few areas of the full parameter space that seem of particular interest and present the results mainly in a set of tables for ease of comparison. The Fisher matrix formalism, if reliably implemented, is far more efficient at probing large parts of the parameter space than full Markov chain Monte Carlo simulations.

We also examine the effect of allowing the endpoint of inspiral to change for spinning objects as has been suggested by Hanna {\it et al.} \cite{Hanna:2008um}. Previous studies have conservatively adopted the Schwarzschild ISCO as the endpoint of the inspiral. The ISCO frequency of an extremal Kerr black hole is roughly a factor seven larger than the ISCO frequency of a Schwarzschild black hole with the same mass. This may allow us to extend the range of validity of the PN waveform and we investigate what effect this has on the parameter estimation.

We focus on determining the intrinsic parameters of mass and spin. Parameters of the source objects such as the distance to the source and the orientation of the binary will affect primarily the amplitude and we will ignore these affects here, effectively by normalising the amplitude to obtain a given signal-to-noise ratio (SNR) and assuming the binary is directly above the interferometer, without any Doppler shift relative to the detector. Here we normalise the SNR to be 10 for a single detector. This is slightly above the threshold for detection and compatible with previous methodology. This choice is merely a normalization of the gravitational wave amplitude and results for other SNR values can be found by a simple linear scaling. The SNR for which the Fisher matrix formalism agrees with Monte Carlo simulations may lie above 10 \cite{Balasubramanian:1995bm}, but this is the conventional scale at which to compare results. Estimates for the performance of multiple detectors operating together can be indicated by combining the individual SNRs in quadrature \cite{gr-qc/9402014}. Multiple detectors have added advantages in dealing with realistic non-Gaussian non-stationary noise, which we do not treat here.

If the compact objects have spin axes that are not aligned with the orbital angular momentum the axes will precess leading to a modulation of the amplitude and other effects \cite{Apostolatos:1994mx}. For simplicity we will also ignore this here. 2PN waveforms, without spin1spin1 self-spin coupling terms, have been investigated for the LISA space detector for the fully aligned case in \cite{Berti:2004bd}, precessing with random spin orientations in \cite{gr-qc/0608062}  and for the partially aligned case in \cite{arXiv:1101.3591}. Further parameters of the source such as mass and energy accretion rates, mass multipole distributions and internal equations of state will also affect the gravitational waveform. The mass multipole correction for spinning neutron stars is second order in the Post-Newtonian approximation \cite{gr-qc/9402014}, but we leave a further investigation of these effects to future work.

\section{Measurement error}

We follow the standard Fisher information methodology of \cite{gr-qc/9402014} and \cite{gr-qc/9502040}. We focus on the case of a single detector, with stationary, non-Doppler shifted, overhead source, for simplicity. With the assumption that the noise in the detector is stationary Gaussian with correlation function $C_{n}(\tau)$, for time separation $\tau$, the probability that the data takes some particular, discretely sampled, pure noise value, $s_{i}$, given an assumption that no signal is present is
\beq p(s_{i}|0) \sim \exp\left(-\frac{1}{2}\frac{s_{i}^{2}}{C_{n}(0)}\right) .\eeq
In the continuum limit, using the Wiener-Kinchin theorem to relate the one-sided power spectral density, $S_{n}(f)$, to the correlation functions, $C_{n}(\tau)$, this can be written as
\beq p(s|0) \sim \exp\left(-2\int^{\infty}_{0}\d f\frac{s(f)s^{*}(f)}{S_{n}(f)}\right) ,\eeq
In practice we will assume that the noise is dominant below some lower frequency bound $f_{_{i}}$ and we will not follow the inspiral waveform beyond an upper frequency $f_{_{f}}$, so we can define an effective inner product such that
\beq \label{innerprod} ( g,h ) \equiv 2\int^{f_{_{f}}}_{f_{_{i}}}\frac{g(f)h^{*}(f)+g^{*}(f)h(f)}{S_{n}(f)}\d f.\eeq
Now we can ask what is the probability of inferring a set of parameters, $\theta$, giving rise to a signal, $h(\theta)$, from a measured detector output, $s$, if there is a true signal $h_{0}$ in the data, such that $s=h_{0}+n$, with $n$ the noise. To do this we use the relation $p(s|h(\theta)) = p(s-h(\theta)|0)$ and ask what is the probability of obtaining an output $s-h(\theta)$, if only noise is present. A generic gravitational waveform strain amplitude, $h(\theta)$, can be expanded around a given reference waveform, $h_{0}$, by
\beq \label{hexpansion} h(\theta) = h_{0} + \Delta\theta_{i}\frac{\partial h}{\partial \theta_{i}} + \frac{1}{2}\Delta\theta_{j}\Delta\theta_{k}\frac{\partial^{2}h}{\partial\theta_{j}\partial\theta_{k}} + ... \eeq
where $\Delta\theta_{i}$ denotes the difference between the $i$th parameter and the value it takes in $h_{0}$. Then the likelihood, $p(s|\theta)$, is given in stationary Gaussian noise by
\bea p(s|\theta) & \sim & e^{-\langle s-h(\theta),s-h(\theta)\rangle/2} \nonumber \\ & = & e^{\left(-\langle n,n\rangle /2 + \Delta\theta_{j}\langle n, h_{j}\rangle - \Delta\theta_{j}\Delta\theta_{k}\langle h_{j},h_{k}\rangle/2 +...\right)} ,\eea
with the assumption that the data, s, is given by $s=h_{0}+n$ and $h_{i}$ is defined as $h_{i} = \partial h/\partial\theta^{i}$. In the Linear Signal Approximation (LSA) all higher derivatives of $h$ can be dropped, and thus terms in the ellipses can be neglected. The term with $\langle n, n \rangle$ is just an overall normalisation and for noise with zero mean the term $\langle n, h_{j} \rangle$ vanishes, so applying the LSA, corresponding to the high Signal to Noise Ratio (SNR) \cite{Vallisneri:2007ev}, the likelihood is just
\beq p(s|\theta) \sim e^{-\theta_{j}\theta_{k}( h_{j},h_{k})/2} ,\eeq
and the posterior probability density function for the parameters $\theta^{i}$ can be found from $p(\theta)p(s|\theta)$. The expectation of the (co)variance of the parameters, in an $n$ parameter model, can be found from the inverse of the Fisher-information matrix
\beq \label{posterior} \langle \Delta\theta_{i}\Delta\theta_{j}\rangle \equiv \frac{\int\Delta\theta_{i}\Delta\theta_{j}p(\theta)p(s|\theta)\d^{n}\theta}{\int p(\theta)p(s|\theta)\d^{n}\theta} ,\eeq
where the angled brackets on the left hand side denote expectation. In the limit of trivial priors $p(\theta)$, that are flat over the entire real line, this becomes
\beq \langle \Delta\theta_{i}\Delta\theta_{j}\rangle = (h_{i},h_{j})^{-1} .\eeq
The Fisher information matrix is generated from the given PN waveforms using the above inner product by
\beq \Gamma_{ij} = \left(\frac{\partial h}{\partial \theta^{i}}\bigg|\frac{\partial h}{\partial \theta^{j}} \right) .\eeq
There are several ways of including the effect of priors on the posterior distribution. Most easily treated are the Gaussian priors used by Cutler and Flanagan \cite{gr-qc/9402014} and Poisson and Will \cite{gr-qc/9502040}. These can be incorporated simply by adding a prior matrix to the Fisher matrix with terms given by $1/\sigma^{2}_{\theta}$, with $\sigma^{2}_{\theta}$ the desired prior variance on the parameter $\theta$. In this case the Gaussian priors are centred around the ``true'' parameter value, so a Gaussian prior imposing a variance of $\sigma^{2}_{\chi} = 1$ on a spin parameter $\chi$, with ``true'' value $\chi=0.95$ will be a Gaussian prior around the value $0.95$ and will allow a non-zero probability of finding a spin greater than $1$. These are called normal true-parameter centered priors (NTC) in \cite{Vallisneri:2007ev}.

Slightly harder to treat, and not employed in the earlier works \cite{gr-qc/9402014} and \cite{gr-qc/9502040} are flat priors bounded by some region, that for example could correspond to positive mass conditions or maximal spin \cite{Vallisneri:2007ev}. In this case care must be taken in choosing suitable parameters with which to impose flat priors. A flat prior on $\mu$, with upper cutoff $\mu_{max}$ is different to a flat prior imposed on $\ln(\mu)$ with upper cutoff $\ln(\mu_{max})$. Also, flat priors in spin parameters $\beta$ and $\sigma$ will not necessarily be the same as flat priors in $\chi_{1}$ and $\chi_{2}$, when the relation between them is non-linear.

The frequencies in (\ref{innerprod}) are the frequencies of the gravitational waves not the orbital frequencies of the massive objects. We will use the approximation that the gravitational waves have a frequency twice the orbital frequency of the objects. This usage is consistent with standard usage in the literature. In the following we will follow standard practice and mainly take $f_{_{i}}=10$Hz and $f_{_{f}}$ as twice the frequency of a test-mass orbiting a Schwarzschild black hole of mass equal to the total mass of our system at the Innermost Stable Circular Orbit (ISCO), that is $f_{_{f}} = 1/(6^{3/2}\pi M)$.

In order to solve (\ref{posterior}) we take note of the fact that the integral, $I$, of a multi-dimensional Gaussian, with respect to a particular parameter $\theta_{p}$ gives error functions
\bea I & = & \int e^{-\Gamma_{_{IJ}} \Delta\theta_{_{I}}\Delta\theta_{_{J}}/2}\d\theta_{p} \nonumber \\ 
& = & e^{-\Gamma_{ij} \Delta\theta_{i}\Delta\theta_{j}/2} \int e^{-\Gamma_{pp} \Delta\theta_{p}^{2}/2 - \Gamma_{pi} \Delta\theta_{p}\Delta\theta_{i}}\d(\Delta\theta_{p}) \nonumber \\ 
& = & \sqrt{\frac{2\pi}{\Gamma_{pp}}}e^{-\left(\Gamma_{ij} - \Gamma_{pi}\Gamma_{pj}/\Gamma_{pp}\right)\Delta\theta_{i}\Delta\theta_{j}/2}\left(\mathrm{Erf}[F(\Delta\theta_{p}^{max})]-\mathrm{Erf}[F(\Delta\theta_{p}^{min})]\right) ,\eea
where $I, J$ are index labels containing the particular value $p$ and $i,j$ are index labels that do not. The argument of the error function in this case is given by 
\beq F[x] = \sqrt{\frac{\Gamma_{pp}}{2}}\left(x+\frac{\Gamma_{pi}}{\Gamma_{pp}}\Delta\theta_{i}\right) .\eeq
When these integration limits, $\Delta\theta_{p}^{max}$ and $\Delta\theta_{p}^{min}$, are taken to $\pm\infty$ the difference in the error functions just evaluates to two and the overall effect of the integration is to project out the direction $\Gamma_{pi}$ in the Fisher metric $\Gamma_{ij}$, up to an overall factor which cancels out in the posterior \cite{Owen:1995tm}.

Multiple parameters can be marginalised in this way. Projecting out all parameters except one leaves just the variance that can be obtained directly by inverting the Fisher matrix. The effect of a flat prior will effectively be to change the limits of integration from $\pm\infty$ to whatever maximum and minimum value the prior allows. In this way flat priors can be included, although multiple flat priors cannot be integrated analytically \cite{Vallisneri:2007ev}.

\section{2.5PN Spinning Waveform}

The induced strain amplitude on the interferometer over time is related to the strain amplitude of the gravitational wave $h(t)$. In order to compute Fisher matrix elements we need the Fourier transform of $h(t)$. The Fourier transform is most efficiently performed analytically using the Stationary Phase Approximation (SPA), which will be valid when the amplitude is changing slowly relative to the frequency of the wave \cite{gr-qc/9402014}. The waveform in the Stationary Phase Approximation is
\beq \tilde{h} = Af^{-7/6}e^{i\Psi}.
\eeq
Schematically the phase can be written as
\beq \Psi = 2\pi f t_{c} + \Psi_{0} + \frac{\Psi_{1}}{f^{5/3}} + \frac{\Psi_{2}}{f} + \frac{\Psi_{3}}{f^{2/3}} + \frac{\Psi_{4}}{f^{1/3}} + \Psi_{5}\log (f) ,\eeq
where $M$ is the total mass. The new term appearing at 2.5PN order is $\Psi_{5}$, although there are also new terms appearing in $\Psi_{0}$ and differences between the $\Psi_{4}$ computed by Arun {\it et al.} \cite{arXiv:0810.5336} and the one used by Poisson and Will \cite{gr-qc/9502040} due to self-spin interaction effects for the two spinning bodies. These self-spin corrections to the 2PN phase were first computed in \cite{Mikoczi:2005dn}. The form of these expansion coefficients, as given in Arun {\it et al.} \cite{arXiv:0810.5336} are;
\beq \Psi_{0} = -\phi_{c}-\frac{\pi}{4} + \frac{\Psi_{5}}{3}\left( 1+\log\left(\frac{\pi{\cal{M}}^{5/2}}{\mu^{3/2}}\right)\right) ,\eeq
\beq \Psi_{1} = \frac{3}{128(\pi{\cal{M}})^{5/3}} ,\eeq
\beq \Psi_{2} = \frac{3715}{32256\pi\mu} + \frac{55\mu^{3/2}}{384\pi{\cal{M}}^{5/2}} ,\eeq
\beq \Psi_{3} = \frac{3{\cal{M}}^{5/6}\beta}{32\pi^{2/3}\mu^{3/2}} - \frac{3\pi^{1/3}{\cal{M}}^{5/6}}{8\mu^{3/2}},\eeq
\beq \Psi_{4} = \frac{15{\cal{M}}^{5/3}}{64\pi^{1/3}\mu^{2}}\left(\frac{3058673}{1016064} + \frac{5429}{1008}\left(\frac{\mu}{{\cal{M}}}\right)^{5/2}  + \frac{617}{144}\left(\frac{\mu}{{\cal{M}}}\right)^{5} - \sigma \right) ,\eeq
\beq \Psi_{5} = \frac{3}{128}\left(\frac{{\cal{M}}}{\mu}\right)^{5/2}\left(\frac{38645\pi}{756}-\frac{65\pi}{9}\left(\frac{\mu}{{\cal{M}}}\right)^{5/2} - \gamma\right) ,\eeq
where,
\bea \beta & = & \frac{113}{24}\sqrt{1-4\left(\frac{\mu}{{\cal{M}}}\right)^{5/2}}(\chi_{1}-\chi_{2}) \nonumber \\ & & + \left(\frac{113}{24}- \frac{19}{6}\left(\frac{\mu}{{\cal{M}}}\right)^{5/2}\right)(\chi_{1}+\chi_{2}) ,\eea
\bea \label{sigma} \sigma & = & \frac{79}{8}\left(\frac{\mu}{{\cal{M}}}\right)^{5/2}\chi_{1}\chi_{2}  + \frac{81}{32}\left(1-2\left(\frac{\mu}{{\cal{M}}}\right)^{5/2}\right)(\chi_{1}^{2}+\chi_{2}^{2}) \nonumber \\ & & + \frac{81}{32}\sqrt{1-4\left(\frac{\mu}{{\cal{M}}}\right)^{5/2}}(\chi_{1}^{2}-\chi_{2}^{2}) ,\eea
\bea \gamma & = & \left(\frac{732985}{4536} - \frac{24260}{162}\left(\frac{\mu}{{\cal{M}}}\right)^{5/2} - \frac{340}{18}\left(\frac{\mu}{{\cal{M}}}\right)^{5}\right)(\chi_{1}+\chi_{2}) \nonumber \\ & & + \left(\frac{732985}{4536}+\frac{140}{18}\left(\frac{\mu}{{\cal{M}}}\right)^{5/2}\right)\sqrt{1-4\left(\frac{\mu}{{\cal{M}}}\right)^{5/2}}(\chi_{1}-\chi_{2}) ,\eea
\beq \mu = \frac{M_{1}M_{2}}{M_{1}+M_{2}} ,\eeq
\beq {\cal{M}} = \frac{(M_{1}M_{2})^{3/5}}{(M_{1}+M_{2})^{1/5}} .\eeq

These are components of a Taylor F2 waveform in the classification of \cite{Damour:2000zb}. $\Psi_{0}$ contains both the phase at coalescence, the term $-\pi/4$ coming from the SPA and spin terms coming from 2.5PN order. Spin terms also occur in the last three terms $\Psi_{3}, \Psi_{4}$ and $\Psi_{5}$. $\Psi_{5}$ only contains spin-orbit contributions in the waveform of \cite{arXiv:0810.5336}. With seven terms with different functional dependencies on $f$ one might expect to be able to reasonably determine seven parameters (including the time at coalescence coming from the linear $f$ term.) To what extent this expectation is borne out depends on the details of the functional dependence of the chirp times and on the precise $f$ dependence through the inner product (\ref{innerprod}). If the parameters are chosen such that the waveform depends linearly on them then the resulting Fisher matrix elements will be independent of the true parameter values. This however, is not the standard practice which we follow here.

Cutler and Flanagan \cite{gr-qc/9402014} considered only spin-orbit contributions at 1.5PN order. Poisson and Will \cite{gr-qc/9502040} considered spin-orbit and spin1-spin2 contributions at 2PN order, only the first term of (\ref{sigma}). In addition, using the waveform of Arun {\it et al.} \cite{arXiv:0810.5336} we consider spin1-spin1, spin2-spin2 terms at 2PN and spin-orbit terms at 2.5PN. In the following, unless stated otherwise, ``2.5PN'' refers to the waveform of \cite{arXiv:0810.5336} and ``2PN'' refers to the waveform used in \cite{gr-qc/9502040}.

To get some impression of the relative weight of the different expansion coefficients we present in Tables \ref{101000}-\ref{530.950} their numerical values for a range of different parameters. We use units where the mass is given in seconds, so that $M_{\odot}=4.93\times 10^{-6}$ seconds and frequencies are measured in Hertz. The coefficients have been rounded to three significant figures. The number of cycles refers to the total number of cycles the gravitational wave goes through between $f_{_{i}}$ and $f_{_{f}}$. As expected the values of $\Psi_{1}$, $\Psi_{2}$ and $\Psi_{3}$ are the same for both the 2.5PN and 2PN cases. In all cases the values of $\Psi_{1}$, $\Psi_{2}$ and $\Psi_{3}$ are the same as the 2PN versions and $\Psi_{4}$ takes the same value for zero spin. There is a difference in the total number of cycles of the gravitational wave of up to 10 cycles. In some cases, for fairly large frequencies, $f\sim 100$Hz, the $\Psi_{5}\log(f)$ term gives a larger contribution to the phase than lower terms such as $\Psi_{1}f^{-5/3}$.

\begin{table}
\begin{tabular}{|c|c|c|c|c|c|c|c|} 
\hline
& $\Psi_{0}$ & $\Psi_{1}$ & $\Psi_{2}$ & $\Psi_{3}$ & $\Psi_{4}$ & $\Psi_{5}$ & no. cycles \\ \hline
2PN & -0.785 & 66200 & 1950 & -1030 & 64.1 & 0 & 227.6 \\ \hline
2.5PN & -104 & 66200 & 1950 & -1030 & 64.1 & 14.5 & 220.4 \\
\hline
\end{tabular}
\caption{Comparison of 2PN and 2.5PN waveform coefficients for the parameter values $M_{1} = M_{2} = 10M_{\odot}$, $\chi_{1}=\chi_{2}=0$ and $\phi_{c}=0$ ($f_{_{ISCO}}=220$Hz).}
\label{101000}
\bigskip
\end{table}

\begin{table}
\begin{tabular}{|c|c|c|c|c|c|c|c|}
\hline
& $\Psi_{0}$ & $\Psi_{1}$ & $\Psi_{2}$ & $\Psi_{3}$ & $\Psi_{4}$ & $\Psi_{5}$ & no. cycles \\ \hline
2PN & -0.785 & 47500 & 1610 & -925 & 61.0 & 0 & 161.6 \\ \hline
2.5PN & -104.9 & 47500 & 1610 & -925 & 61.0 & 15.2 & 154.7 \\
\hline
\end{tabular}
\caption{Comparison of 2PN and 2.5PN waveform coefficients for the parameter values $M_{1} = 15M_{\odot}$, $M_{2} = 10M_{\odot}$ and $\chi_{1}=\chi_{2}=0$ ($f_{_{ISCO}}=176$Hz).}
\bigskip
\end{table}

\begin{table}
\begin{tabular}{|c|c|c|c|c|c|c|c|}
\hline
& $\Psi_{0}$ & $\Psi_{1}$ & $\Psi_{2}$ & $\Psi_{3}$ & $\Psi_{4}$ & $\Psi_{5}$ & no. cycles \\ \hline
2PN & -0.785 & 47500 & 1610 & -384 & 33.8 & 0 & 176.2 \\ \hline
2.5PN & 50.7 & 47500 & 1610 & -384 & 2.98 & -7.5 & 178.2 \\
\hline
\end{tabular}
\caption{Comparison of 2PN and 2.5PN waveform coefficients for the parameter values $M_{1} = 15M_{\odot}$, $M_{2} = 10M_{\odot}$ and $\chi_{1}=0.95$, $\chi_{2}=0.9$ ($f_{_{ISCO}}=176$Hz).}
\bigskip
\end{table}

\begin{table}
\begin{tabular}{|c|c|c|c|c|c|c|c|}
\hline
& $\Psi_{0}$ & $\Psi_{1}$ & $\Psi_{2}$ & $\Psi_{3}$ & $\Psi_{4}$ & $\Psi_{5}$ & no. cycles \\ \hline
2PN & -0.785 & 325000 & 5130 & -2020 & 90.5 & 0 & 1134.2 \\ \hline
2.5PN & -125 & 325000 & 5130 & -2020 & 90.5 & 15.5 & 1124.3 \\
\hline
\end{tabular}
\caption{Comparison of 2PN and 2.5PN waveform coefficients for the parameter values $M_{1} = 5M_{\odot}$, $M_{2} = 3M_{\odot}$ and $\chi_{1}=\chi_{2}=0$ ($f_{_{ISCO}}=550$Hz).}
\bigskip
\end{table}

\begin{table}
\begin{tabular}{|c|c|c|c|c|c|c|c|}
\hline
& $\Psi_{0}$ & $\Psi_{1}$ & $\Psi_{2}$ & $\Psi_{3}$ & $\Psi_{4}$ & $\Psi_{5}$ & no. cycles \\ \hline
2PN & -0.785 & 325000 & 5130 & -1240 & 90.5 & 0 & 1159.3 \\ \hline
2.5PN & 1.39 & 325000 & 5130 & -1240 & 54.6 & -0.27 & 1157.5 \\
\hline
\end{tabular}
\caption{Comparison of 2PN and 2.5PN waveform coefficients for the parameter values $M_{1} = 5M_{\odot}$, $M_{2} = 3M_{\odot}$ and $\chi_{1}=0.95$, $\chi_{2}=0$ ($f_{_{ISCO}}=550$Hz).}
\label{530.950}
\bigskip
\end{table}

\section{Sensitivity curves}

The expected sensitivity of uncompleted interferometers can be simulated. These simulated sensitivites are typically obtained by modelling the thermal noise which dominates at lower frequencies and photon shot noise which dominates at higher frequencies. Seismic noise can be included by adding a sharp cutoff at a low frequency. The initial LIGO interferometers were able to attain their design sensitivity over large parts of the frequency range \cite{LIGO}. With the same waveform and a fixed signal to noise ratio, the shape of the sensitivity curve determines the parameter errors. The power spectral density (PSD) used by Cutler and Flanagan \cite{gr-qc/9402014}, with $f_{0}=70$Hz, is
\beq \label{SfCutFlan} S_{h}(f) = 3\times 10^{-48}\left[\left(\frac{f}{f_{0}}\right)^{-4} + 2\left(1+\left(\frac{f}{f_{0}}\right)^{2}\right)\right] .\eeq
The sensitivity curve used here for advanced LIGO \cite{Cutler:2002me}, with $f_{0}=215$Hz, is
\beq \label{SfadvLIGO} S_{h}(f) = 10^{-49}\left[\left(\frac{f}{f_{0}}\right)^{-4.14} -5\left(\frac{f}{f_{0}}\right)^{-2}+111\left(\frac{1-\left(\frac{f}{f_{0}}\right)^{2}+0.5\left(\frac{f}{f_{0}}\right)^{4}}{1+0.5\left(\frac{f}{f_{0}}\right)^{2}}\right)\right] .\eeq
The curve used by Vallisneri \cite{Vallisneri:2007ev}, which is the initial LIGO curve from Table IV of \cite{Damour:2000zb}, with $f_{0}=150$Hz, is
\beq \label{SfInitLIGO} S_{h}(f) = 9\times 10^{-46}\left[ \left( 4.49\left(\frac{f}{f_{0}}\right)\right)^{-56} + 0.16\left(\frac{f}{f_{0}}\right)^{-4.52}+0.52 +0.32\left(\frac{f}{f_{0}}\right)^{2}\right] .\eeq	
A projected curve for the advanced Virgo detector can be obtained from \cite{Virgo}. 

\section{Results}

We choose to adopt parameters for the waveform $t_{c}$, $\phi_{c}$, $\mu$, ${\cal{M}}$, $\chi_{1}$ and $\chi_{2}$, with $t_{c}$ measured in milliseconds. The choice of these parameters is largely dictated by simplicity and convention. Other combinations of these parameters, such as the symmetric mass ratio, $\eta$, or the spin functions $\beta$ and $\sigma$, as used in \cite{gr-qc/9502040}, can be solved for using the propagation of errors formula \cite{gr-qc/0608062}. The linear spin terms $\beta$, $\sigma$ and $\gamma$ are not independent parameters at 2.5PN order and for our purposes it is easier to use $\chi_{1}$ and $\chi_{2}$ directly.

The waveforms do not distinguish the two black holes if their masses and spins are equal. If we take parameters to be ($t_{c}, \phi_{c}, \ln\mu, \ln{\cal{M}}, \chi_{1}, \chi_{2}$) then the Fisher matrix inversion encounters problems when $M_{1}=M_{2}$ and $\chi_{1}=\chi_{2}$ as the two objects cannot be distinguished and the parameters $\chi_{1}$ and $\chi_{2}$ are degenerate leading to a Fisher matrix with zero determinant, so we deliberately avoid this limit by taking slightly different mass values.

Using the same parameters as \cite{gr-qc/9502040} with $\beta$ and $\sigma$ and their 2PN waveform we are able to reproduce their results up to $\sim 3\%$ accuracy. It is not clear what this discrepancy is due to. The numerical integration has been checked for convergence and reproduces successfully the results of \cite{gr-qc/9402014} using the 1.5PN waveform. The main source of error in our results in the numerical inversion of the Fisher matrix, although for most of our results this is accurate to one part in $10^{7}$. We are also able to reproduce the 2PN results of \cite{Vallisneri:2007ev} to the accuracy reported there using the sensitivity curve (\ref{SfInitLIGO}) and a lower frequency cutoff at $f_{_{i}}=40$Hz and the results for the 2.5PN non-spinning waveforms of \cite{Arun:2004hn} for a lower cutoff at $f_{_{i}}=20$Hz.

\subsection{Differences between 2PN and 2.5PN}

We begin by comparing the 2.5PN waveform of \cite{arXiv:0810.5336} with its seven expansion coefficients and the 2PN waveform used in \cite{gr-qc/9502040} with only six expansion coefficients and slightly different functional dependence in $\Psi_{0}$ and $\Psi_{4}$. In the following, we will denote the waveform used in \cite{gr-qc/9502040} by ``2PN'' and the symbol $\Delta\theta$ is used to denote the root mean squared error relative to the mean - the standard deviation. This is not the same as the difference to the mean used in Eqn. (\ref{hexpansion}), but is consistent with standard notation in the literature.

\begin{table}
\bigskip
\begin{tabular}{|c|c|c|c|c|c|c|}
\hline
& ${\Delta}t_{c}$ & ${\Delta}\phi_{c}$ & $\Delta\mu/\mu$ & $\Delta{\cal{M}}/{\cal{M}}$ & ${\Delta}\chi_{1}$ & ${\Delta}\chi_{2}$ \\ \hline
 & & & & & & \\
2.5PN & 37.0 & 191.5 & 7.74 & 0.0975 & 261.1 & 430.4 \\
 & & & & & & \\
\hline
 & & & & & & \\
2PN  & 22.4 & 248.3 & 21.1 & 0.157 & 14.0 & 44.5 \\
 & & & & & & \\
\hline
\end{tabular}
\caption{Comparison of parameter errors for 2PN and 2.5PN waveforms with parameter values $M_{1}=15M_{\odot}$, $M_{2}=10M_{\odot}$, $\chi_{1}=0.95$ and $\chi_{2}=0.9$.}
\label{15100.950.9noprior}
\bigskip
\end{table}

Parameter error values for true parameters $M_{1}=15M_{\odot}$, $M_{2}=10M_{\odot}$, $\chi_{1}=0.95$ and $\chi_{2}=0.9$, without any priors are given in Table \ref{15100.950.9noprior}. The error on $\mu$ is considerably better for the 2.5PN waveform with a good improvement in ${\cal{M}}$ too. The worsening in the spin parameters is extreme, although neither the 2.5PN nor 2PN waveforms are close to the physically expected bound $|\chi|<1$.

\begin{table}
\bigskip
\begin{tabular}{|c|c|c|c|c|c|c|}
\hline
& ${\Delta}t_{c}$ & ${\Delta}\phi_{c}$ & $\Delta\mu/\mu$ & $\Delta{\cal{M}}/{\cal{M}}$ & ${\Delta}\chi_{1}$ & ${\Delta}\chi_{2}$ \\ \hline
 & & & & & & \\
2.5PN & 7.44 & 13.6 & 0.61 & 0.0050 & 0.77 & 0.92 \\
 & & & & & & \\
\hline
 & & & & & & \\
2PN  & 1.66 & 16.8 & 2.06 & 0.0019 & 0.98 & 1.00 \\
 & & & & & & \\
\hline
\end{tabular}
\caption{Comparison of parameter errors for 2PN and 2.5PN waveforms with parameter values $M_{1}=15M_{\odot}$, $M_{2}=10M_{\odot}$, $\chi_{1}=0.95$ and $\chi_{2}=0.9$ and Gaussian priors on both the spins $\sigma_{\chi_{1}}=1$ and $\sigma_{\chi_{2}}=1$.}
\label{15100.950.9spinprior}
\bigskip
\end{table}

For the same true parameter values with Gaussian priors on both the spins $\sigma_{\chi_{1}}=1$ and $\sigma_{\chi_{2}}=1$, the error values are given in Table \ref{15100.950.9spinprior}. There is still an approximately factor three improvement in $\ln\mu$, but a loss of accuracy in $\ln({\cal{M}})$.

\begin{table}
\bigskip
\begin{tabular}{|c|c|c|c|c|c|c|}
\hline
& ${\Delta}t_{c}$ & ${\Delta}\phi_{c}$ & $\Delta\mu/\mu$ & $\Delta{\cal{M}}/{\cal{M}}$ & ${\Delta}\chi_{1}$ & ${\Delta}\chi_{2}$ \\ \hline
 & & & & & & \\
2.5PN & 3.70 & 23.3 & 0.69 & 0.0039 & 1.4 & 5.0 \\
 & & & & & & \\
\hline
 & & & & & & \\
2PN  & 2.84 & 54.2 & 1.04 & 0.0048 & 0.85 & 10.4 \\
 & & & & & & \\
\hline
\end{tabular}
\caption{Comparison of parameter errors for 2PN and 2.5PN waveforms with parameter values $M_{1} = 5M_{\odot}$, $M_{2} = 1.4M_{\odot}$, $\chi_{1}=0.95$ and $\chi_{2}=0$}
\label{51.40.950noprior}
\bigskip
\end{table}

In Table \ref{51.40.950noprior} we give error values for example parameter values of a light, rapidly spinning black hole and non-spinning neutron star system, with $M_{1} = 5M_{\odot}$, $M_{2} = 1.4M_{\odot}$, $\chi_{1}=0.95$ and $\chi_{2}=0$ without priors. The errors are once again better at 2.5PN for the mass parameters and worse for the $\chi_{1}$ spin parameter, although this time better for $\chi_{2}$. The error difference is still significant in $\mu$ even for these low masses.

\begin{table}
\bigskip
\begin{tabular}{|c|c|c|c|c|c|c|}
\hline
& ${\Delta}t_{c}$ & ${\Delta}\phi_{c}$ & $\Delta\mu/\mu$ & $\Delta{\cal{M}}/{\cal{M}}$ & ${\Delta}\chi_{1}$ & ${\Delta}\chi_{2}$ \\ \hline
 & & & & & & \\
2.5PN & 3.07 & 19.1 & 0.54 & 0.0028 & 0.60 & 0.96 \\
 & & & & & & \\
\hline
 & & & & & & \\
2PN  & 0.79 & 5.42 & 0.22 & 0.0016 & 0.18 & 0.99 \\
 & & & & & & \\
\hline
\end{tabular}
\caption{Comparison of parameter errors for 2PN and 2.5PN waveforms with parameter values $M_{1} = 5M_{\odot}$, $M_{2} = 1.4M_{\odot}$, $\chi_{1}=0.95$ and $\chi_{2}=0$ and Gaussian priors on both the spins $\sigma_{\chi_{1}}=1$ and $\sigma_{\chi_{2}}=1$.}
\label{51.40.950spinspriors}
\bigskip
\end{table}

Errors for the same true parameters with Gaussian priors on both the spins $\sigma_{\chi_{1}}=1$ and $\sigma_{\chi_{2}}=1$ are given in Table \ref{51.40.950spinspriors}. There is still some difference in the parameter estimation errors. From this we conclude that using 2.5PN waveforms of \cite{arXiv:0810.5336} will give noticeably different results to using just the 2PN waveform of \cite{gr-qc/9502040}. It is worth remembering that these results are based on the assumption that the true waveform is a 2.5PN waveform for the 2.5PN numbers and a 2PN waveform for the 2PN ones, neither of which is likely to be true of the full General Relativistic waveform.

\begin{table}
\bigskip
\begin{tabular}{|c|c|c|c|c|c|c|}
\hline
& ${\Delta}t_{c}$ & ${\Delta}\phi_{c}$ & $\Delta\mu/\mu$ & $\Delta{\cal{M}}/{\cal{M}}$ & ${\Delta}\chi_{1}$ & ${\Delta}\chi_{2}$ \\ \hline
 & & & & & & \\
2.5PN & 37.0 & 191.5 & 7.74 & 0.098 & 261.1 & 430.4 \\
 & & & & & & \\
\hline
 & & & & & & \\
2PN with self-spin & 22.4 & 763.8 & 21.1 & 0.157 & 502.4 & 841.0 \\
 & & & & & & \\
\hline
 & & & & & & \\
2PN & 22.4 & 248.3 & 21.1 & 0.157 & 14.0 & 44.5 \\
 & & & & & & \\
\hline
\end{tabular}
\caption{Comparison of parameter values for parameters $M_{1}=15M_{\odot}$, $M_{2}=10M_{\odot}$, $\chi_{1}=0.95$ and $\chi_{2}=0.9$ and $\Psi_{5}=0$. For clarity, ``2PN'' should be here understood as the 2PN waveform used in \cite{gr-qc/9502040}. ``2PN with self-spin'' is 2PN with the spin-orbit and spin-spin terms considered in \cite{gr-qc/9502040} and also the self-spin terms of \cite{arXiv:0810.5336}, which is equivalent to the 2.5PN waveform with $\Psi_{5}=0$.}
\label{15100.950.9Psi50}
\bigskip
\end{table}

To see what causes the large errors differences obtained from the 2PN and 2.5PN waveforms used here, we can artificially remove the $\log(\pi Mf)$ dependent term in the 2.5PN waveform while keeping the self-spin modifications to $\Psi_{4}$. The resulting errors for a rapidly spinning black hole binary system with true parameters $M_{1}=15M_{\odot}$, $M_{2}=10M_{\odot}$, $\chi_{1}=0.95$ and $\chi_{2}=0.9$ are displayed in Table \ref{15100.950.9Psi50}. The errors in the $\chi_{1}$ and $\chi_{2}$ parameters are much larger, larger even than at the full 2.5PN level and there is also a corresponding increase in the error in $\phi_{c}$. This suggests that much of the difference between the 2.5PN waveform of \cite{arXiv:0810.5336} and the 2PN of \cite{gr-qc/9502040} is due to the self-spin interaction terms at 2PN not included in \cite{gr-qc/9502040}.

\begin{table}
\bigskip
\begin{tabular}{|c|c|c|c|c|c|c|}
\hline
& ${\Delta}t_{c}$ & ${\Delta}\Psi_{0}$ & $\Delta\mu/\mu$ & $\Delta{\cal{M}}/{\cal{M}}$ & ${\Delta}\chi_{1}$ & ${\Delta}\chi_{2}$ \\ \hline
 & & & & & & \\
full 2.5PN & 37.0 & 1202.4 & 7.74 & 0.098 & 261.1 & 430.5 \\
 & & & & & & \\
\hline
 & & & & & & \\
2PN with self-spin & 22.4 & 248.3 & 21.1 & 0.157 & 502.4 & 841.0 \\
 & & & & & & \\
\hline
\end{tabular}
\caption{Comparison of errors for parameters $M_{1}=15M_{\odot}$, $M_{2}=10M_{\odot}$, $\chi_{1}=0.95$ and $\chi_{2}=0.9$ and using $\Psi_{0}$ as a parameter, instead of $\phi_{c}$.}
\label{15100.950.9Psi0param}
\bigskip
\end{table}

The larger error ranges noted for the spins may also be partly due to the $\chi_{1}$ and $\chi_{2}$ dependent terms that appear in $\Psi_{0}$ at 2.5PN level and cause confusion with the $\phi_{c}$ term. If we use $\Psi_{0}$ as a parameter, rather than $\phi_{c}$ we obtain results displayed in Table \ref{15100.950.9Psi0param}. The errors for parameters other than $\phi_{c}/Psi_{0}$ are seen to be unchanged. We conclude that mass and spin parameter errors cannot be changed by using $\Psi_{0}$ as a parameter instead of $\phi_{c}$. This is consistent with the findings in \cite{Arun:2004hn}. 

There is still a large difference from the results obtained with the 2PN waveform of \cite{gr-qc/9502040}. The Fisher matrix is more degenerate in the spins when the self-spin couplings are included, meaning it is harder to distinguish the spin values. This large difference in the errors may be partly due to the technical fact that the extra self-spin terms are quadratic in the spin parameters and the Fisher formalism works best with the Linear Signal Approximation \cite{Vallisneri:2007ev}. However, the second derivatives of the waveform phase with respect to the spins was found to be small with respect to the first order terms (by a factor $\sim 0.2$) and the expansion terminates at second order. The waveform phase is also not linear in the chirp mass or symmetric mass ratio, even at the 2PN order considered in \cite{gr-qc/9502040}.

The effect of using large, near extremal true spin values can be compared to assuming their true value is zero. For $M_{1}=15M_{\odot}$, $M_{2}=10M_{\odot}$, $\chi_{1}=\chi_{2}=0$
we find the the errors in $\ln\mu$ and $\ln{\cal{M}}$ increase to 17.13 and 0.14 respectively, but the errors in $\chi_{1}$ and $\chi_{2}$ are only 59.1 and 49.7. This is much lower than the 261.1 and 430.4 found when they are assumed to be nearly maximally spinning. Even when the true values of the spins are zero, including the errors due to the spins has a large detrimental effect on our ability to measure the masses exactly. In the case of $M_{1}=15M_{\odot}$, $M_{2}=10M_{\odot}$, $\chi_{1}=\chi_{2}=0$, if we assume that the spins are exactly zero, and do not marginalise over them then the measured errors in $\ln\mu$ and $\ln{\cal{M}}$ are only 0.020 and 0.0031, an improvement of roughly three, respectively two, orders of magnitude. However, this increase in accuracy will require outside information from other observations or prior restrictions and cannot be justified from the gravitational waveform alone.

\subsection{Effect of frequency ranges and priors}

The final state of the merger of two compact objects is likely to be a spinning black hole. For progenitors with large, aligned spins, the final black hole is likely to have a large spin itself. In this case it is questionable to what extent the ISCO of a non-spinning Schwarzschild black hole is a suitable indication of where the PN inspiral approximation becomes unreliable. A near-extremal Kerr black hole has an ISCO at $r\sim M$, as opposed to $r\sim 6M$ for a non-spinning hole. A spinning black hole can support test particles in quasi-circular obits at much higher frequencies than a non-spinning black hole of the same mass. The asymptotic time period of a test particle in a circular orbit in the equatorial plane of a Kerr black hole is $T=(2\pi(a+\sqrt{r^{3}/M}))$ and hence the gravitational wave frequency at extremality is approximately $f_{_{f}}=1/(2\pi M)$.

There is still a lot of uncertainty about where PN approximations break down and the Schwarzschild ISCO has traditionally been used as a conservative approximation. It is possible \cite{Hanna:2008um} that the PN waveform for a rapidly spinning system is valid to a significantly higher frequency than this. We can look at the approximate effect of increasing the final frequency by allowing $f_{i}$ to increase by a factor of ten. The frequency difference between the true Schwarzschild ISCO and the true extremal Kerr ISCO is closer to a factor of 7, but as there is still uncertainty in exactly where the true limit should be applied, we adopt here a factor of ten for simplicity.

Prior information can also reduce the uncertainty in parameters by a significant amount. In \cite{gr-qc/9402014} and \cite{gr-qc/9502040} Gaussian priors were imposed to simulate in a simple analytical fashion the effect of exact physical priors on the spins of compact objects. To identify the effect of different priors and changing the integration range, in Table \ref{530.950.9variouspriors}, with true parameters $M_{1}=5M_{\odot}$, $M_{2}=3M_{\odot}$, $\chi_{1}=0.95$, $\chi_{2}=0.9$ and the SNR normalised to 10, we implement several different choices of Gaussian spin priors $\sigma_{\chi_{1,2}} = 1$ and a Gaussian coalescence phase prior $\sigma_{\phi_{c}} = \pi$. A longer inspiral waveform with SNR normaised to 10, will have a lower amplitude than a shorter waveform in the corresponding overlap region. 

\begin{table}
\bigskip
\begin{tabular}{|c|c|c|c|c|c|c|}
\hline
& ${\Delta}t_{c}$ & ${\Delta}\phi_{c}$ & $\Delta\mu/\mu$ & $\Delta{\cal{M}}/{\cal{M}}$ & ${\Delta}\chi_{1}$ & ${\Delta}\chi_{2}$ \\
\hline
 & & & & & & \\
no priors & 4.9 & 13.0 & 1.31 & 0.0076 & 17.2 & 31.1 \\
 & & & & & & \\
\hline
 & & & & & & \\
no priors & 1.6 & 9.3 & 0.64 & 0.0043 & 10.9 & 20.0 \\
10$f_{_{ISCO}}$
 & & & & & & \\
\hline
 & & & & & & \\
spin priors & 2.6 & 12.4 & 0.35 & 0.00094 & 0.55 & 0.89 \\
 & & & & & & \\
\hline
 & & & & & & \\
spin priors & 1.1 & 5.7 & 0.18 & 0.00050 & 0.50 & 0.88 \\
10$f_{_{ISCO}}$ 
 & & & & & & \\
\hline
 & & & & & & \\
$\phi_{c}$ prior & 4.2 & 3.1 & 1.26 & 0.0076 & 17.2 & 31.1 \\
 & & & & & & \\
\hline
 & & & & & & \\
spin and $\phi_{c}$ prior & 0.68 & 3.0 & 0.10 & 0.00044 & 0.49 & 0.88 \\
 & & & & & & \\
\hline
 & & & & & & \\
spin and $\phi_{c}$ prior  & 0.56 & 2.8 & 0.099 & 0.00040 & 0.49 & 0.87 \\
10$f_{_{ISCO}}$ & & & & & & \\
\hline
\end{tabular}
\caption{Effect of various combinations of the Gaussian priors $\sigma_{\chi_{1,2}} = 1$ and $\sigma_{\phi_{c}} = \pi$, and increasing the final frequencies for the parameters $M_{1}=5M_{\odot}$, $M_{2}=3M_{\odot}$, $\chi_{1}=0.95$, $\chi_{2}=0.9$ and SNR normalised to 10.}
\label{530.950.9variouspriors}
\bigskip
\end{table}

An increase in $f_{_{f}}$ by a factor $10$ such as might occur in the proposal of \cite{Hanna:2008um}, can lead to a substantial gain in accuracy for unpriored parameters, although the effect diminishes when priors are applied. The Gaussian priors $\sigma_{\phi_{c}} = \pi$ and $\sigma_{\chi_{1,2}} = 1$ contain nearly all the information about these parameters for integrating both to the Schwarzschild ISCO and ten times the Schwarzschild ISCO. The spin priors have a greater effect on both $\Delta\ln\mu$ and ${\Delta}\ln{\cal{M}}$ than the $\phi_{c}$ but the combination of both improves the mass estimation by about a factor twenty. A prior on $\phi_{c}$ was rejected in \cite{Vallisneri:2007ev} since the coalescence phase, $\phi_{c}$, is able to absorb the phase shifts due to changes in the other parameters.

Another prior one could impose is a prior on the masses. The maximum possible value of the symmetric mass ratio $\eta=\mu/(M_{1}+M_{2})$ is $0.25$ which occurs for equal masses. This constraint is simply related to the requirement that the inspiralling objects have real positive masses. It can be translated into a maximum value on $\mu$ of $M/4$ for a fixed total mass $M$. Such a prior can be implemented as a flat prior in the allowed range using the above mentioned likelihood function. The lower bound on the reduced mass at $\mu=0$ is automatically implemented by using the parameter $\ln\mu$ with no bound below.

Imposing the upper bound on $\ln\mu$ it was found that the variance on the $\chi_{1}$ parameter fell from $261.1$ to only $260.5$ for the parameters $M_{1}=15M_{\odot}$, $M_{2}=10M_{\odot}$, $\chi_{1}=0.95$ and $\chi_{2}=0.9$. This is still far from giving a meaningful constraint and cannot exclude the unphysical regime $\chi_{1}>1$. However, the variance in this case does not give equal probabilities on either side of the mode, but is heavily skewed towards lower values and the mode itself is shifted away from the ``true`` value. A flat prior in $\ln\mu$ corresponds to a logarithmic prior in the parameter $\mu$ since it implies that each decade of $\mu$ has equal prior probability.

A more impressive constraint can be obtained by using the parameter $\eta$ directly instead of $\ln\mu$. In this case the cutoff at $\eta=0.25$ can be imposed directly as a flat prior in $\eta$ between 0 and 0.25. For high masses $M_{1}=15M_{\odot}$, $M_{2}=10M_{\odot}$, and near extremal spins $\chi_{1}=0.95$ and $\chi_{2}=0.9$, $\Delta\chi_{1}$ is reduced from 261.1 to 27.6 and $\Delta\chi_{2}$ is reduced from 430.4 to 47.3 (There is also a small shift in the Fisher matrix values, presumably due to the non-linear change of parameters from $\ln\mu$ to $\eta$.) For lower masses, $M_{1}=5M_{\odot}$, $M_{2}=3M_{\odot}$, with true spins set to zero, the reduction in $\Delta\chi_{1}$ is from 22.9 to 6.9 and the reduction in $\Delta\chi_{2}$ is from 34.9 to 12.5. For the unequal mass BH-NS case, $M_{1}=10M_{\odot}$, $M_{2}=1.4M_{\odot}$ with $\chi_{1}=0.95$ and $\chi_{2}=0$, $\Delta\chi_{1}$ reduces from 1.9 to 1.2 and $\Delta\chi_{2}$ from 15.5 to 10.6. These reductions are still not sufficient to place the spins inside the physical bound and there does not seem to be any strong reason for preferring a flat prior in $\eta$ to a flat prior in $\ln\mu$ other than the more direct relation to the cutoff at $\eta=0.25$.

Although it does not seem possible with the 2.5PN waveform to assign well measured spin values to the individual components of the binary, one of the main reasons for this is that the spin parameters $\chi_{1}$ and $\chi_{2}$ are not very well distinguished by the 2.5PN aligned spins waveform. The two spin parameters are highly degenerate. However, the error ellipse given by the likelihood function obtained by marginalising over all parameters except the spins is often very thin in one direction. This means that there is a particular linear combination of the spin parameters that is much better constrained by the waveform than the individual spins. In \cite{Ajith:2009bn} it was suggested that this combination might be
\beq \chi = \frac{1+\delta}{2}\chi_{1} + \frac{1-\delta}{2}\chi_{2} ,\eeq
where $\delta=(M_{1}-M_{2})/(M_{1}+M_{2})$. In the case with $M_{1}=15M_{\odot}$, $M_{2}=10M_{\odot}$ we have $\delta = 0.2$ and thus $\chi = 0.6\chi_{1}+0.4\chi_{2}$.

With a two-dimensional likelihood function, obtained by marginalising over the other parameters, the preferred direction can be found by finding the eigenvectors of the associated Hessian matrix. Numerically with $M_{1}=15M_{\odot}$, $M_{2}=10M_{\odot}$, $\chi_{1}=0.95$ and $\chi_{2}=0.9$ we find $\chi = 0.86\chi_{1}+0.52\chi_{2}$. The error on this linear combination is $\Delta\chi = 0.91$ and the maximum value it can attain, when both spins are at their maximal physical value, $\chi_{1}=\chi_{2}=1$, is 1.37. For the same high masses, but $\chi_{1}=\chi_{2}=0$ we find $\chi=0.60\chi_{1} + 0.80\chi_{2}$. The error on this combination is $\Delta\chi=33.2$ and the maximum value it can attain is 1.4.

For lower masses, such as $M_{1}=5$, $M_{2}=3$ with near maximal, aligned spins, $\chi_{1}=0.95$ and $\chi_{2}=0.9$, we find $\chi=0.88\chi_{1}+0.48\chi_{2}$ as the optimally measured linear combination of spin parameters. The error on this combination is just $\Delta\chi = 0.31$ and its maximum possible physical value is 1.36. With no true spin we obtain instead  $\chi=0.84\chi_{1}+0.55\chi_{2}$ as the optimally measured linear combination. This combination has a measurement error of $\Delta\chi = 1.28$. The maximum value that $\chi$ can attain if both spins are at their maximal physical value is 1.38. This suggests that in certain parameter ranges, particularly lower masses, such as neutron stars or light black holes, it may be possible to demonstrate that the system contains a non-zero spin within a physical bound, although separate spins cannot easily be assigned to the individual objects.

In the case that extra external information is able to precisely fix the spins of one of the objects, perhaps the lighter object, then the parameter estimation problem becomes essentially five dimensional. This might happen if one of the objects is a pulsar, or a neutron star with known equation of state that prevents it from having any meaningful spin angular momentum. In this case the spin of the heavier object can be measured much more accurately without the need for priors (although the spin of the other object is essentially a delta function prior). The low mass case $M_{1}=5$, $M_{2}=3$ with true parameters $\chi_{1}=0.95$ and $\chi_{2}=0$ are compared in Table \ref{530.9506vs5params} and a black hole-neutron star example, with $M_{1}=10$, $M_{2}=1.4$ and true parameter $\chi_{1}=0.95$ and $\chi_{2}=0$ in Table \ref{15100.9506vs5params}. In the case of a low mass companion, with the lighter spin known exactly, it may be possible to determine that the spin of the heavier object lies within the physical range to $1\sigma$, without the use of any priors.

\begin{table}
\bigskip
\begin{tabular}{|c|c|c|c|c|c|c|}
\hline
& ${\Delta}t_{c}$ & ${\Delta}\phi_{c}$ & $\Delta\mu/\mu$ & $\Delta{\cal{M}}/{\cal{M}}$ & ${\Delta}\chi_{1}$ & ${\Delta}\chi_{2}$ \\ \hline
 & & & & & & \\
6 params & 4.7 & 28.7 & 1.52 & 0.0082 & 21.4 & 30.6 \\
$\chi_{2}$ unknown & & & & & & \\
\hline
 & & & & & & \\
5 params  & 2.9 & 19.6 & 0.50 & 0.0016 & 1.85 & 0.0 (exactly) \\
$\chi_{2}$ known & & & & & & \\
\hline
\end{tabular}
\caption{Comparison of estimation of six parameters versus five parameters for true parameter values $M_{1}=5$, $M_{2}=3$, $\chi_{1}=0.95$ and $\chi_{2}=0$.}
\label{530.9506vs5params}
\bigskip
\end{table}

\begin{table}
\bigskip
\begin{tabular}{|c|c|c|c|c|c|c|}
\hline
& ${\Delta}t_{c}$ & ${\Delta}\phi_{c}$ & $\Delta\mu/\mu$ & $\Delta{\cal{M}}/{\cal{M}}$ & ${\Delta}\chi_{1}$ & ${\Delta}\chi_{2}$ \\ \hline
 & & & & & & \\
6 params & 7.0 & 15.0 & 0.84 & 0.0079 & 1.9 & 15.5 \\
$\chi_{2}$ unknown & & & & & & \\
\hline
 & & & & & & \\
5 params  & 4.7 & 13.2 & 0.38 & 0.0025 & 0.28 & 0.0 (exactly) \\
$\chi_{2}$ known & & & & & & \\
\hline
\end{tabular}
\caption{Comparison of estimation of six parameters versus five parameters for true parameter values $M_{1}=10$, $M_{2}=1.4$, $\chi_{1}=0.95$ and $\chi_{2}=0$.}
\label{15100.9506vs5params}
\bigskip
\end{table}

We also present a comparison of the 2.5PN inspiral waveform and the phenom waveform (which includes 2.5PN spin terms) from \cite{arXiv:1005.3306} in Table \ref{PNvsPhenom}. This is necessarily only a four-dimensional parameter set, as the way spins are treated is slightly different in the two cases. Effectively here we are assuming that the true spins are zero and known to be exactly zero.

With the merger and ringdown modeled we can extend the integration range well beyond the ISCO frequency, here we can take it to $0.15/M$ (suggested by Frank Ohme) which corresponds to 1218Hz instead of 176Hz for the Schwarzschild ISCO with total mass $25M_{\odot}$ and 3800Hz instead of 550Hz for the total mass $8M_{\odot}$ case. When extending the frequency cutoff, to ensure a fair comparison, we can fix the amplitude and allow the SNR to increase. In the high mass example, there is some increase in the SNR from 10 to 11.72 but a slight deterioration in the mass parameters' accuracy. In the low mass example the SNR only increases from 10 to 10.0147 but there is an increase in error on $\ln\mu$ and no meaningful change in the chirp mass.

\begin{table}
\bigskip
\begin{tabular}{|c|c|c|c|c|}
\hline
& ${\Delta}t_{c}$ & ${\Delta}\phi_{c}$ & $\Delta\mu/\mu$ & $\Delta{\cal{M}}/{\cal{M}}$ \\ \hline
$M_{1}=15M_{\odot}$, $M_{2}=10M_{\odot}$ & & & &  \\
2.5PN waveform & 1.8 & 3.04 & 0.020 & 0.0031  \\
$\chi_{1}=\chi_{2}=0$ & & & &  \\
\hline
$M_{1}=15M_{\odot}$, $M_{2}=10M_{\odot}$ & & & &  \\
phenom waveform & 2.3 & 3.69 & 0.027 & 0.0039  \\
$\chi=0$, $f_{_{f}}=$ ISCO & & & &  \\
\hline
$M_{1}=15M_{\odot}$, $M_{2}=10M_{\odot}$ & & & &  \\
phenom waveform & 1.6 & 2.83 & 0.022 & 0.0035  \\
$\chi=0$, $f_{_{f}}=0.15/M$& & & &  \\
\hline
$M_{1}=5M_{\odot}$, $M_{2}=3M_{\odot}$ & & & &  \\
2.5PN waveform & 0.56 & 1.60 & 0.0070 & 0.00037  \\
$\chi_{1}=\chi_{2}=0$ & & & &  \\
\hline
$M_{1}=5M_{\odot}$, $M_{2}=3M_{\odot}$ & & & &  \\
phenom waveform & 0.51 & 1.34 & 0.0059 & 0.00036  \\
$\chi=0$, $f_{_{f}}=$ ISCO & & & &  \\
\hline
$M_{1}=5M_{\odot}$, $M_{2}=3M_{\odot}$ & & & &  \\
phenom waveform & 0.41 & 1.15 & 0.0052 & 0.00033  \\
$\chi=0$, $f_{_{f}}=0.15/M$& & & &  \\
\hline
\end{tabular}
\caption{Comparison of PN inspiral waveform and phenom waveform of \cite{arXiv:1005.3306} for various true parameters and integration ranges.}
\label{PNvsPhenom}
\bigskip
\end{table}

This analysis has ignored the spins, effectively assuming that they are known and zero, but it does not appear that there is much to gain from using this particular phenom waveform, even after the frequency range is greatly extended and in this way the SNR somewhat increased for the same amplitude signal. This interesting area though deserves more study as better phenom and hybrid Post-Newtonian and numerical waveforms become available.

\subsection{Shape of sensitivity curve}

It is also interesting to compare the effect of different shapes of sensitivity curve. The sensitivity curves used in \cite{gr-qc/9502040}, \cite{Vallisneri:2007ev} and here are all different, see Eqns (\ref{SfCutFlan}-\ref{SfInitLIGO}). Using the low mass parameters $M_{1}=5M_{\odot}$, $M_{2}=3M_{\odot}$, $\chi_{1}=0.95$, $\chi_{2}=0$, with a normalised SNR$=10$, an initial frequency $f_{_{i}}=10$Hz and $f_{_{f}}$ at the Schwarzschild ISCO we obtain the results displayed in Table \ref{sensecurves}.

\begin{table}
\bigskip
\begin{tabular}{|c|c|c|c|c|c|c|}
\hline
& ${\Delta}t_{c}$ & ${\Delta}\phi_{c}$ & ${\Delta}\mu/\mu$ & $\Delta{\cal{M}}/{\cal{M}}$ & ${\Delta}\chi_{1}$ & ${\Delta}\chi_{2}$ \\ \hline
 & & & & & & \\
SfadvLIGO & 4.7 & 28.7 & 1.5 & 0.0082 & 21.4 & 30.6 \\
no priors & & & & & & \\
\hline
 & & & & & & \\
SfadvLIGO & 1.63 & 10.6 & 0.26 & 0.00086 & 0.89 & 0.95 \\
spin priors & & & & & & \\
\hline
 & & & & & & \\
SfInitLIGO & 12.2 & 38.7 & 8.5 & 0.074 & 184.6 & 294.1 \\
no priors & & & & & & \\
\hline
 & & & & & & \\
SfInitLIGO & 1.5 & 10.5 & 0.28 & 0.0019 & 0.97 & 0.99 \\
spin priors & & & & & & \\
\hline
 & & & & & & \\
SfCutFlan & 6.9 & 42.7 & 1.45 & 0.0066 & 14.5 & 18.6 \\
no prior & & & & & & \\
\hline
 & & & & & & \\
SfCutFlan & 2.3 & 12.8 & 0.28 & 0.0010 & 0.95 & 0.97 \\
spin priors & & & & & & \\
\hline
 & & & & & & \\
SfadvVirgo & 3.6 & 23.1 & 1.16 & 0.0061 & 15.7 & 22.2 \\
no prior & & & & & & \\
\hline
 & & & & & & \\
SfadvVirgo & 1.4 & 9.9 & 0.24 & 0.00079 & 0.86 & 0.93 \\
spin priors & & & & & & \\
\hline
\end{tabular}
\caption{Comparison of different sensitivity curves both with and without Gaussian spin priors for true parameters $M_{1}=5M_{\odot}$, $M_{2}=3M_{\odot}$, $\chi_{1}=0.95$, $\chi_{2}=0$.}
\label{sensecurves}
\bigskip
\end{table}

We see that the advanced Virgo curve appears to be the best both with and without priors. This is likely due to the wider, but flatter shape of the advanced Virgo sensitivity curve. These sensitivities refer to singular detectors operating in isolation. When operating together with combined analyses the results will be better, partly through an improvement in effective SNR \cite{gr-qc/9402014}.

Finally, we list some results from various different regions of the parameter space in Table \ref{variousparams}. With both spins anti-aligned to the orbital angular momentum the spin uncertainty becomes very large.

\begin{table}
\bigskip
\begin{tabular}{|c|c|c|c|c|c|c|}
\hline
& ${\Delta}t_{c}$ & ${\Delta}\phi_{c}$ & ${\Delta}\mu/\mu$ & $\Delta{\cal{M}}/{\cal{M}}$ & ${\Delta}\chi_{1}$ & ${\Delta}\chi_{2}$ \\ \hline
$5M_{\odot}$, $3M_{\odot}$ & & & & & & \\
$\chi_{1}=0$, $\chi_{2}=0$ & 4.3 & 80.0 & 1.80 & 0.0089 & 22.9 & 34.9 \\
no priors & & & & & & \\
\hline
$5M_{\odot}$, $3M_{\odot}$ & & & & & & \\
$\chi_{1}=0.95$, $\chi_{2}=0.9$ & 4.9 & 13.0 & 1.3 & 0.0076 & 17.2 & 31.1 \\
no priors & & & & & & \\
\hline
$5M_{\odot}$, $3M_{\odot}$ & & & & & & \\
$\chi_{1}=0.95$, $\chi_{2}=-0.9$ & 4.3 & 55.1 & 1.73 & 0.0087 & 24.9 & 27.0 \\
no priors & & & & & & \\
\hline
$5M_{\odot}$, $3M_{\odot}$ & & & & & & \\
$\chi_{1}=0$, $\chi_{2}=0$ & 0.6 & 9.6 & 0.35 & 0.0024 & 0.86 & 0.94 \\
spin priors & & & & & & \\
\hline
$15M_{\odot}$, $10M_{\odot}$ & & & & & & \\
$\chi_{1}=0.95$, $\chi_{2}=0.9$ & 37.0 & 191.5 & 7.7 & 0.097 & 261.1 & 430.4 \\
no priors & & & & & & \\
\hline
$15M_{\odot}$, $10M_{\odot}$ & & & & & & \\
$\chi_{1}=0$, $\chi_{2}=0$ & 27.0 & 536.6 & 17.1 & 0.14 & 59.1 & 49.7 \\
no priors & & & & & & \\
\hline
$15M_{\odot}$, $10M_{\odot}$ & & & & & & \\
$\chi_{1}=0.95$, $\chi_{2}=-0.9$ & 26.8 & 373.8 & 17.3 & 0.14 & 607.9 & 745.6 \\
no priors & & & & & & \\
\hline
$15M_{\odot}$, $10M_{\odot}$ & & & & & & \\
$\chi_{1}=-0.95$, $\chi_{2}=0.9$ & 18.9 & 1160 & 24.9 & 0.17 & 991.4 & 1051.7 \\
no priors & & & & & & \\
\hline
$15M_{\odot}$, $10M_{\odot}$ & & & & & & \\
$\chi_{1}=-0.95$, $\chi_{2}=-0.9$ & 21.1 & 1623.7 & 23.1 & 0.17 & 994.0 & 1398.4 \\
no priors & & & & & & \\
\hline
$15M_{\odot}$, $10M_{\odot}$ & & & & & & \\
$\chi_{1}=0.95$, $\chi_{2}=-0.9$ & 2.2 & 6.5 & 0.13 & 0.0033 & 0.996 & 0.997 \\
spin priors & & & & & & \\
\hline
$10M_{\odot}$, $1.4M_{\odot}$ & & & & & & \\
$\chi_{1}=0.95$, $\chi_{2}=0$ & 7.0 & 15.0 & 0.84 & 0.0079 & 1.9 & 15.5 \\
no priors & & & & & & \\
\hline
$10M_{\odot}$, $1.4M_{\odot}$ & & & & & & \\
$\chi_{1}=-0.95$, $\chi_{2}=0$ & 7.0 & 149.0 & 0.88 & 0.0081 & 11.4 & 82.1 \\
no priors & & & & & & \\
\hline
$10M_{\odot}$, $1.4M_{\odot}$ & & & & & & \\
$\chi_{1}=0$, $\chi_{2}=0$ & 6.5 & 99.3 & 1.0 & 0.0088 & 4.9 & 35.0 \\
no priors & & & & & & \\
\hline

\end{tabular}
\caption{Parameter errors for various different true parameter values for the 2.5PN aligned spinning waveform and advanced LIGO sensitivity curve. Some values are also given with Gaussian priors on the spins of $\sigma_{\chi_{1}}=1$ and $\sigma_{\chi_{2}}=1$.}
\label{variousparams}
\bigskip
\end{table}

\section{Conclusions}

We have obtained a number of results related to parameter estimation using Post-Newtonian 2.5 order spinning waveforms. The most important result is that measurement errors differ significantly from previous results obtain by Poisson and Will \cite{gr-qc/9502040} at 2PN order. A large part of this difference is attributable to the self-spin interaction terms at 2PN order not included in the analysis of \cite{gr-qc/9502040} or in analyses for space-based missions \cite{Berti:2004bd}. However, it also strongly suggests that 2PN spinning waveforms are not sufficient for reliable parameter estimation from spinning systems and possibly 2.5 PN order is not sufficient either. Using the best available 2.5PN waveform leads to a Fisher matrix that is more degenerate in the spins than previous studies. This also means that small changes in the waveform can have large effects on the computed errors. Source multipole moments needed to obtain the spin contributions for gravitational waves have recently been computed at 3PN order \cite{Porto:2010zg} and it is hoped that the results reported here will give extra impetus to putting these contributions into a ``ready to use'' form necessary for data analysis and parameter estimation. Higher order spinning waveforms may also be needed beyond 3PN.

We have also explored a number of issues with the 2.5PN waveform. Implicitly the analysis presented here assumes that nature corresponds exactly to this 2.5 PN order. The model of an exactly aligned, 2.5PN order, spinning waveform of this form is a prior assumption in the analysis. Furthermore the analysis assumes that an individual detector will have a stationary Gaussian noise spectrum given, for example, by the modeled approximation (\ref{SfadvLIGO}) and ignores any other systematic errors in the observational and data manipulation processes. In light of these facts, and the belief that nature is unlikely to be exactly 2.5 PN, it is difficult to give precise estimates for the parameter errors that can be obtained from real advanced second generation interferometers. The values given above should taken as purely indicative of how signals might behave in certain parts of parameter space and which parts of parameter space might be amenable to constraining which parameters. Fisher matrix calculations have been compared with Monte Carlo methods in \cite{Balasubramanian:1995bm} and there it was found that Fisher matrix techniques can in cases underestimate the errors by a factor of two.

We have shown that while the mass parameters are often well constrained using the 2.5PN waveform, allowing in cases black hole candidates to be distinguished from neutron star candidates, the spin information is only weakly constrained. Only in the case where one of the spins is known exactly from external information can the spin of the other object be constrained within the physical bound $|\chi_{1}|<1$.

The detectors have limited sensitivity to intrinsic spin at 2.5PN level and information about the spins is still likely to be dominated by prior information. Certain information that is necessary in the parameter estimation is difficult to fold into the Fisher matrix formalism. More accurate prior information can be encoded into the likelihood using a variety of techniques \cite{Vallisneri:2007ev}. The method used here employs a semi-analytical technique that greatly reduces computational complexity, but is limited to one flat prior and the exact choice of parameter is an important consideration in applying ``flat'' priors. The prior information that may be so important to obtaining reliable mass estimates, is still not well implemented.

Assumptions about where the inspiral PN breaks down, using Kerr ISCO instead of Schwarzschild ISCO, does have some effect, but prior information is still dominant in these cases. Phenom waveforms do not have sufficient information or accuracy yet to completely alleviate this problem. The 2.5PN level expansion is unlikely to be sufficient for reliable extraction of spin parameters in General Relativity. As advanced hybrid waveforms are developed containing a full waveform from inspiral to ringdown in full General Relativity more work will be needed to study this interesting area.

Several proposals for measuring deviations from General Relativity rely on measuring simultaneously the parameters needed to describe the binary using the waveform and also checking for deviations from the predictions of General Relativity. With non-spinning systems the waveforms are largely just a function of the two masses, which can be measured fairly accurately using only a few chirp times, leaving the remaining chirp times as a consistency check of the model predictions. For aligned spinning systems, there are four intrinsic parameters to be measured to describe the system and as we have seen the spins are not well determined, even within the physically expected range. It is possible that waveforms at 3PN and above will alleviate some of these problems encountered here and leave this to further work.

\section{Acknowledgments}

It is a pleasure to thank Curt Cutler, Tom Dent, Badri Krishnan, Andrew Lundgren and Frank Ohme for useful discussions on this work.

\section{Appendix}

We give here for completeness examples of the covariance matrices, $\Sigma^{ij}$. (We thank Drew Keppel for the suggestion that this would be useful to include.) The square root of the diagonal elements corresponds to the measured accuracy of each parameter and the correlation between parameters $a$ and $b$ can be calculated by the formula
\beq
c_{_{ab}}=\frac{\Sigma^{ab}}{\sqrt{\Sigma^{aa}\Sigma^{bb}}}
\eeq
For the case of six parameters, with $t_{c}=\phi_{c}=0$, $M_{1}=15M_{\odot}$, $M_{2}=10M_{\odot}$, $\chi_{1}=0.95$ and $\chi_{2}=0.9$ we have
\beq
\Sigma^{ij}_{6}=
\left(
\begin{array}{cccccc}
 0.00137 & -6.627 & -0.282 & 0.00343 & -9.323 & 15.356 \\
 -6.627 & 36669.7 & 1451.79 & -18.632 & 49757.5 & -82071.1 \\
 -0.282 & 1451.79 & 59.831 & -0.744 & 2008.48 & -3310.16 \\
 0.00343 & -18.632 & -0.744123 & 0.00951 & -25.398 & 41.886 \\
 -9.323 & 49757.5 & 2008.48 & -25.398 & 68152.8 & -112369. \\
 15.356 & -82071.1 & -3310.16 & 41.886 & -112369. & 185276.
\end{array}
\right)
\eeq
With only four parameters, $t_{c}=\phi_{c}=0$, $M_{1}=15M_{\odot}$, $M_{2}=10M_{\odot}$ and the spins assumed to be exactly known at their values $\chi_{1}=0.95$ and $\chi_{2}=0.9$, we have
\beq
\Sigma^{ij}_{4}=
\left(
\begin{array}{cccc}
 5.906\times 10^{-6} & 0.0102 & -0.000107 & 6.272\times 10^{-6} \\
 0.0102 & 18.091 & -0.191 & 0.0115 \\
 -0.000107 & -0.191 & 0.00202 & -0.00012 \\
 6.272\times 10^{-6} & 0.0115 & -0.000124 & 9.141\times 10^{-6}
\end{array}
\right)
\eeq
The corresponding correlations are
\beq
c_{_{ab}}=
\left(
\begin{array}{cccccc}
 1. & -0.935 & -0.986 & 0.950 & -0.965 & 0.964 \\
 -0.935 & 1. & 0.980 & -0.998 & 0.995 & -0.996 \\
 -0.986 & 0.980 & 1. & -0.987 & 0.995 & -0.994 \\
 0.950 & -0.998 & -0.987 & 1. & -0.998 & 0.998 \\
 -0.965 & 0.995 & 0.995 & -0.998 & 1. & -0.999 \\
 0.964 & -0.996 & -0.994 & 0.998 & -0.999 & 1.
\end{array}
\right)
\eeq
\beq
c_{_{ab}}=
\left(
\begin{array}{cccc}
 1. & 0.992 & -0.981 & 0.854 \\
 0.992 & 1. & -0.997 & 0.895 \\
 -0.981 & -0.997 & 1. & -0.915 \\
 0.854 & 0.895 & -0.915 & 1.
\end{array}
\right)
\eeq
The parameters are all highly correlated, especially the spins.

\end{document}